\documentstyle[aps,prl]{revtex}
\begin{document}
\preprint{}
\draft
\title{Theory of a Novel Superstring in Four Dimensions}
\author{B. B. Deo}
\address{Physics Department, Utkal
University, Bhubaneswar-751004, India.} 
\maketitle
\begin{abstract}
An open string in four dimensions is supplemented by
forty four Majorana fermions. The fermions are grouped in
such a way that the resulting action is supersymmetric.
The super-Virasoro algebra is constructed and closed by
the use of Jacobi identity. The tachyonic ground state 
decouples from the physical states. After a GSO projection, 
the resulting physical mass spectrum is shown to be 
$\alpha^{\prime} M_n^2 =n$ where $n=0, 1, 2,\cdots $. There are 
fermions and bosons in each mass level. The internal symmetry 
group of the string breaks to $SU(3) \times SU(2)\times U(1)
\times U(1)$.
\end{abstract}

String theory was invented \cite{bd1} as a sequel to dual
resonance models \cite{bd2} to explain the properties of 
strongly interacting particles in four dimensions. Assuming
the string to live in a background gravitational field and
demanding Weyl invariance, the Einstein equations of general
relativity could be deduced. It was believed that about these
classical solutions one can expand and find the quantum
corrections. But difficulties arose at the quantum level.
Eventhough the strong interaction amplitude obeyed crossing, 
it was no longer unitary. There were anomalies and ghosts. 
Therefore it was necessary for the open string to live in 26
dimensions \cite{bd3}. in this letter we report that 
fermionising the 22 additional coordinates for a four 
dimensional open string and regrouping them, we can obtain 
a novel superstring with an unusual but correct mass spectrum
containing fermions and bosons.

There are forty four Majorana fermions representing the 22 
bosonic coordinates \cite{bd11}. We divide them into four 
groups or `generations' (one time-like and three space-like).
They are labelled by $\mu=0,1,2,3$ like the space time indices
and each contain 11 fermions. These 11 fermions are again
divided into two groups, one containing six and the other five. 
For convenience, in one group we have $j=1,3,5,7,9,11$
six odd numbers, and in other, the five even numbers 
$k=2,4,6,8,10$.

The string action is
\begin{equation}
S=-\frac{1}{2 \pi} \int d^2 \sigma \left [ {\partial}_{\alpha} 
X^{\mu}{\partial}^{\alpha} X_{\mu}-i \bar{\psi}^{\mu,j}
\rho^{\alpha} {\partial}_{\alpha}\psi_{\mu,j}-i\phi^{\mu,k} 
\rho^{\alpha}{\partial}_{\alpha}\phi_{\mu,k}\right ]\;,
\end{equation}
$\rho^{\alpha}$ are the two dimensional Dirac matrices
\begin{equation}
\rho^0=\left ( \begin{array}{cc} 0 & -i\\
i & 0 \end{array} \right )\;,\;\;\;\;\;
\rho^1=\left ( \begin{array}{cc} 0 & i\\
i & 0 \end{array} \right )
\end{equation}
and obey 
\begin{equation}
\{\rho^{\alpha}, \rho^{\beta}\}=-2 \eta_{\alpha \beta}\;.
\end{equation}
In general we follow the notations and conventions of reference
\cite{bd5} whenever omitted by us. $X^{\mu}(\sigma,\tau)$ are 
the string coordinates. The $\psi$'s are the odd indexed and
$\phi$'s the even indexed Majorana fermions decomposed 
in the basis
\begin{equation}
\psi=\left (\begin{array}{c}\psi_- \\ \psi_+ \end{array}
\right )\;,
\;\;\;\mbox{and}\;\;\;\;
\phi=\left (\begin{array}{c}\phi_- \\ \phi_+ \end{array}
\right )\;.
\end{equation}
The nonvanishing commutation and anticommutations are
\begin{equation}
[\dot{X}^{\mu} (\sigma,\tau),X^{\nu}(\sigma^{\prime},\tau)]
=-i\;\delta(\sigma-\sigma^{\prime})\eta^{\mu \nu}
\end{equation}
\begin{equation}
\{\psi_A^{\mu}(\sigma,\tau), \psi_B^{\nu}(\sigma^{\prime},\tau)\}
=\pi\; \eta^{\mu \nu}\; \delta_{AB}\; \delta(\sigma-\sigma^{\prime})\}
\end{equation}
\begin{equation}
\{\phi_A^{\mu}(\sigma,\tau), \phi^{\nu}(\sigma^{\prime}, \tau)
=\pi\; \eta^{\mu \nu}\; \delta_{AB}\; \delta(\sigma-\sigma^{\prime})\}
\end{equation}

The action is invariant under infinitesimal transformations
\begin{equation}
\delta X^{\mu}=\bar \epsilon\; \left ( \sum_j \psi^{\mu,\;j}
+i\sum_k\phi^{\mu,\;k}\right )
\end{equation}
\begin{equation}
\delta\psi^{\mu,\;j}=-i \rho^{\alpha}\; \partial_{\alpha} X^{\mu}
\;\epsilon
\end{equation}
\begin{equation}
\delta \phi^{\mu,\;k}=+\rho^{\alpha}\;\partial_{\alpha}\; X^{\mu}
\;\epsilon
\end{equation}
where $\epsilon$ is an infinitesimally constant anticommuting
Majorana spinor. The commutator of the two supersymmetry
transformation gives a spatial translation, namely
\begin{equation}
[\delta_1,\delta_2]X^{\mu}=a^{\alpha}\;{\partial}_{\alpha}X^{\mu}
\end{equation}
and
\begin{equation}
[\delta_1,\delta_2]\Psi^{\mu}=a^{\alpha}{\partial}_{\alpha}
\Psi^{\mu}
\end{equation}
where 
\begin{equation}
a^{\alpha}=2i\;\bar \epsilon_1\; \rho^{\alpha}\; \epsilon_2
\end{equation}
and
\begin{equation}
\Psi^{\mu}=\sum_j\psi^{\mu,\;j}+i\sum_k\phi^{\mu,\;k}
\end{equation}
In deriving this, the Dirac equation for the spinors have been
used. The Noether super-current is
\begin{equation}
J_{\alpha}=\frac{1}{2} \rho^{\beta}\;\rho_{\alpha}
\;\Psi^{\mu}\; \partial_{\beta} X_{\mu}
\end{equation}

We now follow the standard procedure. The light cone components
of the current and energy momentum tensors are
\begin{equation}
J_+=\partial_+X^{\mu}\; \Psi^{\mu}_+
\end{equation}
\begin{equation}
J_-=\partial_-X^{\mu}\;\Psi^{\mu}_-
\end{equation}
\begin{equation}
T_{++}=\partial_+X^{\mu} \partial_+X_{\mu}+\frac{i}{2}
\psi_+^{\mu,\;j}\;\partial_+\psi_{+\mu,\;j}+\frac{i}{2}
\phi_+^{\mu,\;k}\;\partial_+\phi_{+\mu,\;k}
\end{equation}

\begin{equation}
T_{--}=\partial_-X^{\mu} \partial_-X_{\mu}+\frac{i}{2}
\psi_-^{\mu,\;j}\partial_-\psi_{-\mu,\;j}+\frac{i}{2}
\phi_+^{\mu,\;k}\partial_+\phi_{+\mu,\;k}
\end{equation} 
where $\partial_{\pm}=\frac {1}{2} (\partial_{\tau}\pm 
\partial_{\sigma})$.

The theory is quantised with
\begin{equation}
\partial_{\pm}X^{\mu}=\frac{1}{2}\sum_{_\infty}^{+\infty}
\alpha_n^{\mu}\; e^{-in(\tau\pm\sigma)}
\end{equation}
\begin{equation}
[\alpha_m^{\mu},\alpha_n^{\nu}]=m\; \delta_{m+n}\;
\eta^{\mu \nu}
\end{equation}

In this work, we always choose the Neveu-Schwarz (NS) \cite{bd4}
boundary condition. Then the mode expansions of the fermions are
\begin{equation}
\psi_{\pm}^{\mu,j}(\sigma,\tau)=\frac{1}{\sqrt 2}
\sum_{r\in Z+\frac{1}{2}}b_r^{\mu,\;j}e^{-ir (\tau+\sigma)}
\end{equation}

\begin{equation}
\phi_{\pm}^{\mu,k}(\sigma,\tau)=\frac{1}{\sqrt 2}
\sum_{r\in Z+\frac{1}{2}}d_r^{\mu,\;k}e^{-ir (\tau \pm \sigma)}
\end{equation}
The sum is over all the half-integer modes.
\begin{equation}
\{b_r^{\mu,j}, b_s^{\nu, j^{\prime}}\}=\eta^{\mu \nu}\;\delta_
{j,j^{\prime}}\; \delta_{r+s}
\end{equation}
\begin{equation}
\{d_r^{\mu,k}, b_s^{\nu, k^{\prime}}\}=\eta^{\mu \nu}\;\delta_
{k,k^{\prime}}\;\delta_{r+s}
\end{equation}

Virasoro generators \cite{bd6} are given by the modes of the
energy momentum tensor $T_{++}$ and Noether current $J_+$,
\begin{equation}
L_m^M=\frac{1}{\pi}\int_{-\pi}^{+\pi} d\sigma\; e^{im\sigma}\; T_{++}
\end{equation}
\begin{equation}
G_r=\frac{\sqrt 2}{\pi}\int_{-\pi}^{+\pi} d\sigma\; e^{ir\sigma}\; J_{+}
\end{equation}
$`M'$ stands for matter. In terms of creation and annihilation
operators
\begin{equation}
L_m^M=L_m^{(\alpha)}+L_m^{(b)}+L_m^{(d)}
\end{equation}
where
\begin{equation}
L_m^{(\alpha)}=\frac{1}{2}\sum_{n=-\infty}^{\infty}:
\alpha_{-n}\cdot\alpha_{m+n}:
\end{equation}

\begin{equation}
L_m^{(b)}=\frac{1}{2}\sum_{r=-\infty}^{\infty}
(r+\frac{1}{2}m) :b_{-r}\cdot b_{m+r}:
\end{equation}

\begin{equation}
L_m^{(d)}=\frac{1}{2}\sum_{r=-\infty}^{\infty}
(r+\frac{1}{2}m):d_{-r}\cdot d_{m+r}:
\end{equation}

In each case normal ordering is required. The single dot
implies the sum over all qualifying indices. 
The fermionic generator is
\begin{equation}
G_r=\sum_{n=-\infty}^{\infty} \alpha_{-n}\cdot
(b_{r+n}+i\;d_{r+n})
\end{equation}
The Virasoro algebra in this NS sector is
\begin{equation}
[L_m^M, L_n^M]=(m-n) L_{m+n}^M+A(m)\;\delta_{m+n}
\end{equation}

\begin{equation}
[L_m^M, G_r]=\left (\frac{1}{2}m-r\right ) G_{m+r}
\end{equation}

The anticommutator $\{G_r, G_s\}$ is not directly obtainable
as there are mixed terms. Instead we use the Jacobi identity
\begin{equation}
[\{G_r,G_s\},L_m^M]+\{[L_m^M,G_r],G_s\}+\{
[L_m^M,G_s],G_r\}=0
\end{equation}
which implies, consistent with equations (34) and (35),
\begin{equation}
\{G_r,G_s\}=2 L_{r+s}^M+B(r)\delta_{r+s}
\end{equation}
$A(m)$ and $B(r) $ are normal ordering anomalies. Taking the 
vacuum expectation value in the Fock ground state $|0,0\rangle $ 
with four momentum $ p^{\mu}=0$ of the commutator $[L_1,L_{-1}]$
and $[L_2, L_{-2}]$, it is easily found that
\begin{equation}
A(m)=\frac{26}{8}(m^3-m)
\end{equation}
and using the Jacobi identity for $L_0$, $G_{1/2}$, $G_{-1/2}$ 
and $L_{-2}$, $ G_{3/2}$, $G_{1/2}$ for the Fock ground
state, one obtains
\begin{equation}
B(r)=\frac{26}{2}\left (r^2-\frac{1}{4}\right )
\end{equation}
The central charge $c=26$. This is what is expected.
Each bosonic coordinate contribute 1 and each fermionic
ones contribute $1/2$, so that the total central charge is +26.

A physical state satisfies
\begin{equation}
 L_m^M\;|phys\rangle=0\;\;\;\;\;\;\;\;\;n>0
\end{equation}
For obtaining a zero central charge so that the anomalies
cancel out and also ghosts are isolated, Faddeev-Popov 
(FP) ghosts \cite{bd7} are introduced. The FP ghost action
is
\begin{equation}
S_{FP}=\frac{1}{\pi}\int (c^+\partial_- b_{++} + c^-
\partial_+b_{--})d^2 \sigma
\end{equation}
where the ghost fields $b$ and $c$ satisfy the anticommutator
relations.
\begin{equation}
\{b_{++}(\sigma,\tau), c^+(\sigma^{\prime},\tau)\}
=2 \pi\; \delta(\sigma-\sigma^{\prime})
\end{equation}

\begin{equation}
\{b_{--}(\sigma,\tau), c^-(\sigma^{\prime},\tau)\}
=2 \pi\; \delta(\sigma-\sigma^{\prime})
\end{equation}
and are quantized with the mode expansions
\begin{equation}
c^{\pm}=\sum_{-\infty}^{\infty}c_n\; e^{-in(\tau\pm\sigma)}
\end{equation}

\begin{equation}
b_{\pm \pm}=\sum_{-\infty}^{\infty}b_n\; e^{-in(\tau\pm\sigma)}
\end{equation}
The canonical anticommutator relations for $c_n$'s and
$b_n$'s are
\begin{equation}
\{c_m,b_n\}=\delta_{m+n}
\end{equation}
\begin{equation}
\{c_m,c_n\}=\{b_m,b_n\}=0
\end{equation}

Deriving the energy momentum tensor from the action and making
the mode expansion, the Virasoro generators for the ghosts (G)
are
\begin{equation}
L_m^G=\sum_{n=-\infty}^{\infty}(m-n)\;b_{m+n}\; c_{-n}- a\; \delta_{m}
\end{equation}
where $a$ is the normal ordering constant. These generators 
satisfy the algebra
\begin{equation}
[L_m^G,L_n^G]=(m-n)\;L_{m+n}^G+A^G(m)\; \delta_{m+n}
\end{equation}
The anomaly term is deduced as before and give
\begin{equation}
A^G(m)=\frac{1}{6}(m-13m^3)+2a\;m
\end{equation}
With $a=1$, this anomaly term becomes
\begin{equation}
A^G(m)=-\frac{26}{12}(m^3-m)
\end{equation}
The central charge is $-26$ and cancels the $A(m)$ term of the
$L_m^M$'s.

The BRST \cite{bd8} charge operator is
\begin{equation}
Q_{BRST}=\sum_{-\infty}^{\infty}L_{-m}^M\;c_m -\frac{1}{2}
\sum_{-\infty}^{\infty}(m-n) :c_{-m}\; c_{-n}\; b_{m+n} :
-a\; c_0
\end{equation}
and the nilpotent for $a=1$. The physical states are such
that $Q_{BRST}\;|phys\rangle=0$.

The ghosts are not coupled to the physical states.
Therefore the latter must be of the form (up to null state).
\begin{equation}
|\{n\}\; p\rangle_M \otimes\; c_1|0\rangle_G
\end{equation}
$|\{n\}\; p\rangle_M$ denotes the occupation numbers and momentum of 
the physical matter states. The operator $c_1$ lowers the
energy of the state by one unit and is necessary for BRST
invariance. The ghost excitation is responsible for lowering
the ground state energy which produces the tachyon \cite{bd9}.
\begin{equation}
(L_0^M-1)\;|phys\rangle=0
\end{equation}
Therefore, the mass shell condition is
\begin{equation}
\alpha^{\prime} M^2 = N^B+N^F-1
\end{equation}
where 
\begin{equation}
N^B=\sum_{m=1}^{\infty}\alpha_{-m}\; \alpha_m
\end{equation}
\begin{equation}
N^F=\sum_{r=1/2}^{\infty}r\;(b_{-r}\; b_r+d_{-r}\;d_r)\;.
\end{equation}
$\alpha^{\prime}M^2$ takes the values $-1, -1/2, 0, 1/2, 
1, 3/2, 2$, and so on. The tachyonic ground state
$\alpha^{\prime}M^2=-1$, belongs to the ghost sector. 
The remaining states fall into two categories. Defining 
G-parity as $(-1)^F$ where $F$ is the fermionic number, 
the half integral spin states of the above spectrum have
odd parity. We make the GSO projection by demanding that the
physical states have $G=+1$ i.e., even G-parity \cite{bd10}.
Then the physical state mass spectrum can be described by
\begin{equation}
\alpha^{\prime}M^2=n
\end{equation}
where $n$ takes values $0,1,2,\cdots $. This is the Regge 
conjecture. The zero mass state can be a vector boson or
a pair of fermions.

This supersymmetry is natural with no SUSY particles.
We now discuss the internal symmetry of the group. Three
space-like generations each with $SO(11)$ symmetry are 
clearly discernible. It is well known \cite{bd12} that
$SO(11)\supset SU(5)\times U(1) \times SU_C(3)$ and that 
$SO(5)$ can break to $SU(2) \times U(1)$ by the adjoint
representation. The resulting symmetry group is $SU_C(3)
\times SU(2) \times U(1) \times U(1) $. The additinal
$U(1)$ makes no difference to the predictions of the 
standard model. However there can be another $Z$- boson
which may be observable.

We have presented the complete theory of a novel superstring
in four dimensions. We hope that this string theory will
encompass the standard model on one hand and general theory
of relativity on the other.

The Library, Computer and other facilities extended by Institute
of Physics, Bhubaneswar is thankfully acknowledged.

\end{document}